\documentclass[11pt]{article}
\usepackage{amsmath}
\usepackage{natbib}

\usepackage{color}

\usepackage{graphicx}

\newcommand{\Cov}{\mbox{Cov}}

\date{\today}

\begin{document}

\title{Beyond Homophily:
Incorporating Actor Variables
in Actor-oriented Network Models}

\author{
Tom A.B. Snijders\\
University of Groningen, University of Oxford\\
Department of Sociology,
Grote Rozenstraat 31, 9712TG Groningen, \\The Netherlands
\and
Alessandro Lomi\\
University of Italian Switzerland, Lugano
}

\maketitle

\newpage

{\LARGE\it
\begin{center}
Beyond Homophily: Incorporating Actor Variables \\
in Statistical Network Models
\end{center}
}
\bigskip

\hyphenation{Snij-ders}

\begin{abstract}
We consider the specification of effects of numerical actor attributes,
measured according to an interval level of measurement,
in statistical models for directed social networks.
A fundamental mechanism is homophily or assortativity, where actors have
a higher likelihood
to be tied with others having similar values of the variable under study.
But there are other mechanisms that may also play a role in how
the attribute values of two actors influence the likelihood of a tie between them.
We discuss three additional mechanisms:
aspiration, defined as the tendency to send more ties to others
having high values;
what we call attachment conformity, defined by sending more ties to others
whose values are close to what may be considered the `social norm';
and sociability,
where those having higher values will tend to send more ties generally.
These mechanisms may operate jointly, and then their effects will be confounded.
We present a specification
representing these effects simultaneously by a four-parameter quadratic function
of the values of sender and receiver.
Greater flexibility can be obtained by a five-parameter extension.
We argue that empirical researchers often overlook the possibility that
homophily may be confounded with these other mechanisms, and that for
numerical actor attributes that have important effects on directed networks,
these specifications may provide an improvement.

An illustration is given of the dependence of advice ties
on academic grades in a network of MBA students, analyzed by the
Stochastic Actor-oriented Model.
\medskip

Keywords: actor covariate, directed networks, homophily, assortativity,
aspiration, conformity, sociability,
Stochastic Actor-oriented Model, quadratic model,
academic performance.
\end{abstract}

\section{Introduction and overview}

Representing dependence between tie variables is of
paramount importance for the specification of statistical network models.
This is done by using so-called structural effects.
For the use of such models in empirical research, however,
representing the effects of nodal variables is also essential.
The importance of nodal variables, also known as (monadic) attributes
or actor covariates, was already recognized by
\citet{FienbergWasserman1981}, who showed how to use
categorical attributes in log-linear models for network data.
The ability to combine structural and covariate effects is an important
feature also for more recent network models such as the Exponential
Random Graph Model \citep{WassermanPattison96,ERGMbook},
the Stochastic Actor-oriented Model \citep{Snijders01},
and Latent Space Models \citep{HRH02}.
What precisely is represented by covariate effects in the network literature
has been varying, but the main focus has been on homophily,
the tendency for actors to relate to others who are similar in
terms of a limited number of contextually salient dimensions
\citep{LazarsfeldMerton54,McPhersonEA01,Azoulay2017}.
This is also called assortativity; we use the term `homophily'.
Methodological discussions about how to model homophily mostly focus on
similarity based on binary or other categorical variables.
This has informed many empirical studies.
Examples of binary and categorical variables frequently used to
specify homophily include gender, occupation, and membership
in ethnic, religious, or other social categories \citep{McPherson2004}.
Some examples of continuous and ordinal variables considered for homophily
are age, education, and various attitudinal scales.

However, the importance of nodal variables goes well beyond homophily
effects. In this paper we elaborate this specifically for
statistical models for binary social networks where the set of network ties
constitutes the dependent variable, and for numerical actor covariates
that satisfy an interval level of measurement.
Such variables might be truly continuous, such as length or monetary
variables, but also discrete, such as sum totals of psychological
multi-item scales. The set of actor variables considered also
includes discrete ordinal variables provided that their numerical values
are interpreted as having an interval level of measurement.
Whether this is acceptable is a matter of choice, and depends on the
interpretation in the case at hand. Sometimes even the well-known Likert scale
with values, e.g., 1 to~5, may be interpreted as having approximately
interval-level scale properties. A requirement then is that it is reasonable
to consider the values as being equidistant in how they are interpreted.

As a scope condition for the network, we only consider directed relations
where a tie from sender $i$ to receiver $j$ can be interpreted as the result
of a positive choice, in some sense, originating from $i$ to the target $j$.

In the next section we propose a basic set of four mechanisms
according to which such variables
might affect dyadic probabilities of tie existence, creation, and/or termination.
These are similarity/homophily, attachment conformity, aspiration, and sociability.
Their definitions follow below.
We focus on directed networks because the asymmetry between senders
and receivers of ties permits a clear distinction between these
mechanisms.
These four patterns can be represented jointly
by a quadratic function of the values of
senders and receivers, having a total of four statistical parameters.
This model is proposed in Section~3, followed by a five-parameter extension
that adds flexibility but decreases parsimony.
The model is applied in an example in Section~\ref{S_ex},
in a longitudinal study of an advice relation in an educational setting,
employing the Stochastic Actor-oriented Model.
A discussion section concludes the paper.
The main conclusion is that researchers interested in modeling
actor covariates in statistical network models should go beyond
the automatism of considering only homophily and stopping there,
and rather consider a wider array of covariate effects
that merit consideration, such as the quadratic model proposed here.

\section{Homophily and other principles of attraction}

Homophily, the tendency to have or form ties to others with the same or
similar characteristics, is a dominant principle of
dyadic attraction between social actors \citep{McPhersonEA01},
but it is not the only one.
Discussing attraction in the context of network modeling,
\citet{Stokman2004} \citep[published as][]{StokmanVieth04} distinguishes
three types of dimensions influencing interpersonal attraction:
similarity, aspiration, and complementarity.
What is called similarity-attraction may have several aspects,
depending on which units are being compared for the
assessment of similarity.
The tendency toward \emph{homophily} means that sender and receiver
are compared, and ties become more likely
as their similarity increases along relevant dimensions.
But social actors making choices about sending ties may also
compare potential recipients of the tie with a reference group,
i.e., with what is socially considered appropriate;
a positive tie then will be more likely when the recipient is more
similar to the norm describing what is appropriate
\citep{Sherif1936,Homans74,Cohen77,AbramsEtAl1990}.
We define \emph{attachment conformity} as the tendency
that ties are more likely when the recipient's characteristics are closer
to one particular value, common to all; this value then is called
the `social norm', but we do not go further into its substantive
interpretation because this will depend also on the rest of the model.
To avoid confusion, note that conformity has two faces:
the high value put on others who display
normative behaviour --which is what is treated here--, and the adjustment
of behaviour toward normative values -- part of the same mechanism,
but not considered here.
(Another closely related concept is cumulative advantage or
preferential attachment \citep{Price76,BarAlbert99}, defined as the
tendency to send ties preferably to those who have high degrees already.
This concept is outside the scope of our discussion, because it is
not directly related to an exogenous actor variable; but it is clear that
attachment conformity for actor variables may have consequences
that are similar to preferential attachment.)

An \emph{aspiration} dimension is an attribute for which high values are
generally found attractive\footnote{This definition is different from Stokman's
(2004) first definition, which is formulated in terms of aspiration to belong to
a group to which one currently does not belong;
but the further interpretation is similar.}.
For a negative aspiration dimension, low values are generally found attractive;
since this is just a mirror image obtainable by changing the sign of the variable,
we discuss only positive aspiration.
This means that the attribute is seen as being positively related to
the quality or competence of the receiving
actor, for purposes that are directly or indirectly associated with
the relation under consideration.
Aspiration is a concept used more generally in psychological
theories of goal setting \citep{LewinEtAl1944,Knudsen2008},
and aspiration dimensions are quite basic in interpersonal attraction.
\citet{Robins2009} coins the word `capacities' for individual
factors, such as skills, expertise, information or knowledge,
that may `bear on social actions' (\emph{op.\ cit.}) and lead
to a higher number of ties for actors commanding them.
Already \citet{LottLott1965} reviewed studies finding that high-status
and warm individuals are more likely to be considered attractive.
Social status is often an important summary signal of quality
\citep{SauderLynnPodolny2012} and, accordingly,
status variables may be expected to often have an aspiration aspect.
\citet{SelfhoutEA10} studied how friendship dynamics
is influenced by personality characteristics,
as defined by the `Big Five' \citep{McCraeJohn1992}.
They found strong evidence for attraction toward
persons high on agreeableness, which is defined as the extent
to which a person is cooperative, kind, and trusting.
The aspirational dimension of social selection becomes especially evident
in studies of status and reputation where differences in social
or material resources controlled by social actors
(directly or through their connections with others) produce
systematic differences in the attractiveness of potential
partners, as can be seen in  \citet{KilKra1994}
and \citet{StuartEA1999}.

Aspiration may be regarded as a boundary case of attachment conformity,
where the social norm corresponds to a very high value of the attribute.
While recognizing this relation, we nevertheless mention them
distinctly; in the mathematical implementation, this issue will reappear.

Complementarity, or heterophily \citep{RSU2010}, is a social selection
mechanism in which relations are more likely to be observed between
actors who have different
attributes, and the combination of attributes is especially valuable.
Complementarity plays a role especially in exclusive dyadic relations
and also at the level of the personal network, where a focal actor may wish
to have a diverse network composition with, e.g.,
\underline{at least one} person who has a desired complementary,
hence different, attribute.
It often involves the combination of several variables \citep{RSU2010}.
We do not consider complementarity further.

The three dimensions of homophily, attachment conformity, and aspiration are
naturally treated as principles of attraction, for actors having
the distinct roles of senders and receivers in a directed network,
the sender being attracted to potential receivers to a lower or
higher degree.
Homophily is about the combination of the characteristics of sender and
receiver, while attachment conformity and aspiration are about the receiver's
characteristic.
A dimension at the side of the sender of ties is \emph{sociability},
also referred to as gregariousness, activity, or outgoingness.
A sociability dimension is a characteristic for which high values
are associated to sending many ties.
Thereby it is a mirror image of aspiration.
Variables indicating high resources will be expected to be
sociability dimensions, because they help overcome the costs
of sending ties. For friendship, \citet{SelfhoutEA10} found that
-- not surprisingly -- extraversion is associated with sociability.




The main point we wish to make in this paper is that ordinal and numerical
actor variables will often have a combination of similarity, attachment conformity,
aspiration, and sociability aspects for choices about dyadic relations.
The next section shows how this may be represented in statistical network
models.
How for a given variable
the combination of similarity, attachment conformity, aspiration, and sociability
works out will depend on the variable, the actor set and its context,
and the relation under consideration.

Some examples of their combinations are the following.
For the role of health-related lifestyle variables such as smoking and
drinking habits in friendship relations, homophily may be the
most important mechanism,
but attachment conformity and aspiration may also play a role.
E.g., in some groups drinking may confer high status and therefore
be associated with aspiration \citep[e.g.,][]{OsgoodEA2013},
whereas in other groups drinking moderately may be
the norm so that drinking habits will be associated with attachment conformity.
Further, individuals who drink more might make more friends -- sociability.

For relations that involve cooperation toward some goal,
e.g., collaboration or advice, it is possible that individuals are prone
to seek contacts with those who have high values on variables signaling good
performance, such as expertise. Therefore performance-signaling attributes
may be associated with the mechanism of aspiration;
but similarity and normative behavior decrease uncertainty and facilitate
cooperation, and therefore the mechanisms of homophily and attachment
conformity could be at play here, too.
An example is \citet{BrouwerEA2018}.

For a wide variety of social relations
high-status others may be desirable interaction partners,
but ties crossing large status gaps might be uneasy to manage or
violate social norms; the former would be in line with aspiration,
the latter with homophily but also with attachment conformity.
\citet{Podolny1994} argues that actors in markets prefer ties
to others with higher status; since this is everybody's preference,
the ties that form will be between actors of similar status.
This is to be expected particularly in social settings that are
hierarchically structured such as formal organizations,
or when status is interpreted a signal for underlying qualities
that are not directly observable \citep{SauderLynnPodolny2012}.

Of course it is an abstraction to focus on only one attribute,
and in real life there will be a multiplicity of attributes at work,
confounded, interacting, and/or endogenously influencing each other.
Homophily interactions for multiple actor variables are discussed
by \citet{BlockGrund2014}; some other recent examples of studies
carefully considering the interplay of multiple actor variables are
\citet{Schaefer2018} and \citet{GremmenEA2018}.
In this paper we focus on modeling a single attribute.
Our ideas can be used also in studies with multiple attributes.

The considerations presented above apply to ordinal variables generally,
and the requirement of numerical variables comes into play only when formulas
are specified for the mathematical specification. The following parts
of the paper are specific for numerical variables with an
interval level of measurement (permitting some give and take
with respect to this requirement, as mentioned in the Introduction).
Dichotomous variables fall outside the scope because for them, attachment conformity
is perfectly confounded with aspiration;
this can be positive or negative aspiration, depending on whether the
social norm corresponds to the higher or the lower category.
For dichotomous actor variables the combination of sender and receiver effects
entails three degrees of freedom, and these are completely represented by
using a sender, receiver, and their interaction
effect (the latter could equivalently be replaced by a `same', `absolute difference' or
`similarity' effect).

\section{Representing effects of actor attributes}

In this section we discuss how homophily, attachment conformity, aspiration, and
sociability can be expressed in statistical models for networks.
The numerical actor attribute that is the variable under consideration
will be denoted by $V$. It is assumed to be one-dimensional.
The value of $V$ for actor $i$ is denoted $v_i$.
The network will be represented by tie variables $x_{ij}$ such
that $x_{ij} = 1$ indicates the presence of a tie $i \rightarrow j$
from sender~$i$ to receiver~$j$, and $x_{ij} = 0$ its absence.

In most statistical network models, the probabilities of ties  ---or
tie changes--- depend on a linear predictor such as used
in generalized linear models.
The linear predictor is a function of the entire network $x$.
The part of the linear predictor depending on $V$
will here be called the \emph{social selection function}, with the caveat
that it does not represent preference or attraction per se:
as is generally the case in statistical models with correlated
variables, it models just probabilities, and the representation of how $V$
influences tie choices is not determined totally by this social selection
function but depends also on other correlated effects.
For linear statistical models we have the machinery of partial
and semi-partial correlations and coefficients to analyze this,
but in network models the dependence structure is more complicated
and partialing approaches have not yet been developed.
Lacking a more precise set of tools,
we shall just keep in mind that the social selection function is
somewhat similar in interpretation, but not identical, to
a preference function.
To avoid cumbersome language, we nevertheless shall sometimes use the
word `attraction' meaning something like `sending a tie to this actor
with a higher probability if all other circumstances are equal'.

A basic issue for the representation of actor attributes in statistical
network models is that this representation links the monadic level of
individual actors to the dyadic level of network ties, illustrating the
fundamental multilevel nature of network analysis \citep{Snijders16}.
For the tie directed from $i$ to $j$, the two actors involved
have values $v_i$ and $v_j$, respectively, for variable $V$.
We only consider social selection functions of the form
\begin{equation}
    \sum_{i,j} \, x_{ij}\, a( v_j \mid v_i) \ , \label{base}
\end{equation}
the summation extending over all network members.
This means that $a( v_j \mid v_i)$ represents the influence of $V$
on the probability of the existence, creation, or maintenance
of the tie $i \rightarrow j$. With a slight misuse of terminology
we shall also refer to $a( v_j \mid v_i)$ as the social selection function.

Given our interpretation of the tie $i \rightarrow j$ as resulting
from a choice by $i$ of actor $j$, we consider the social selection
function $ a( v_j \mid v_i)$ for given $v_i$ as a function of $v_j$,
and explore which families of functions
are useful as social selection functions.
For this purpose, we require that ---depending on the values of
the parameters in this family--- the family of functions can represent
combinations of tendencies toward homophily, toward attachment conformity,
toward aspiration, and toward sociability.
The \emph{optimum} of the function is defined as the highest value
of $a( v_j \mid v_i)$  as a function of $v_j$ for a given $v_i$.

Dyadic attraction as influenced by numerical actor variables,
seen in the perspective of a sender choosing a receiver, is often
formulated in terms of \emph{ideal points} \citep{Coombs1964,Jones1983}.
Given a preference function of an actor, an ideal point is an
argument value where an optimum is assumed; if the preference function
is unimodal, this will be a unique point.
Although the social selection function $a( v_j \mid v_i)$ is not strictly
interpreted as a preference function, the parallel with preference functions
still can teach us some things.
For preference functions depending on actor attributes,
for an attribute leading exclusively to homophily, the ideal point
is the actor's own value; for an attribute representing aspiration
in its strongest sense,
where the attraction to others becomes higher when the attribute
gets larger, it is the highest possible value of the covariate (or infinity);
for an attribute representing pure attachment conformity,
it is the value corresponding to the social norm, common to all actors
and hence independent of $v_i$.
In a suitable family of social selection functions,
depending on its parameters any of these various points should be
possible as the location of its optimum. Further, to represent
sociability, depending on the parameters the function should be
able to be generally increasing in $v_i$ on the whole range of $V$.

\subsection{Usual representations}

Homophily with respect to a numerical actor variable is usually represented
in statistical network models by making the probability
of existence or change of a tie depend on the absolute difference
between $v_i$ and $v_j$,
\begin{equation}
   a_1( v_j \mid v_i) \,=\,  \beta_1 \mid v_i - v_j \mid \ .    \label{simX}
\end{equation}
For $\beta_1 < 0$ this is a function with its optimum in $v_j = v_i$.
This representation tends to be used almost without reflection;
see \citet{Snijders01} and \citet{SnijdersEA10b} for actor-oriented models,
\citet{Goodreau2007} and \citet{ERGMbook} for exponential random graph models,
and, among many other examples, \citet{DSZ2004} and \citet{Louch2000}
for various other models.
This is a very parsimonious representation, requiring only one
parameter, but inflexible because the optimum can only be assumed
in $v_j = v_i$, and therefore this can represent only pure homophily,
not aspiration or attachment conformity.
Sometimes main effects of the sender's and receiver's values are added,
\begin{equation}
   a_1( v_j \mid v_i) \,=\, \beta_1 \, v_i + \beta_2\, v_j +
      \beta_3  \left| v_i - v_j \right| ,   \label{simXplus}
\end{equation}
and then $\beta_1$ is meant to represent sociability,
$\beta_2$ to represent aspiration, and $-\beta_3$ homophily.
The parameters can only be interpreted together, however.
The optimum is assumed in $v_j = v_i$
if and only if $\beta_3 \leq \left| \beta_2 \right|$;
and there is generally aspiration to high values of $v_j$
if and only if $\beta_2 > \left| \beta_3 \right|$.
This means that homophily and aspiration are not readily
combined in this model, and attachment conformity cannot be represented.

As an alternative, sometimes the ego-by-alter product interaction
is proposed to represent homophily \citep[eg.,][]{SnijdersEA10b}.
The main effects of sender and receiver then also have to be included,
leading to
\begin{equation}
   a_2( v_j \mid v_i) \,=\, \beta_1 \, v_i + \beta_2\, v_j + \beta_3\, v_i\,v_j
    \ .  \label{egoXaltX}
\end{equation}
This expresses, for $\beta_3 > 0$, that senders with higher $v_i$
have a higher tendency to connect to receivers with high $v_j$.
Considering (\ref{egoXaltX}) as a function of $v_j$,
for given $v_i$, shows
this function can be linearly decreasing or increasing,
switching this behavior at $v_i = -\beta_2/\beta_3$.
Thus it represents not homophily but differential aspiration;
assuming that $\beta_3 > 0$, attraction is toward low values of $V$
for egos with $v_i < -\beta_2/\beta_3$, and toward high values
for $i$ with $v_i > -\beta_2/\beta_3$.
Thus, the two functions $a_1(v_j \mid v_i)$ and $a_2(v_j \mid v_i)$  have
fundamentally different properties.

This shows that the most commonly used models for representing
effects of numerical actor variables on tie creation and change
are, respectively, a model representing only pure homophily,
a model combining homophily, aspiration, and sociability
in a rather inflexible way,
and another model representing pure differential aspiration.
In practice, however, actor variables may be associated with
homophily, attachment conformity, aspiration, as well as sociability, and any
combination of these mechanisms;
and researchers hardly ever have enough strong theoretical knowledge
to be able to \emph{a priori} exclude some of these possibilities
and make a confident bet on only one or two of them.

\subsection{Quadratic representations}

For a combination of the four mechanisms potentially associated to the
attribute $V$ we need a parametric family of
functions that can represent unimodal as well as monotone functions;
with the property that -- for the unimodal type -- the
location of the optimum can be close to ego's value to represent
homophily, can be drawn toward a common (`normative') value to
represent attachment conformity, and can be higher or lower to represent aspiration.

Absolute differences such as used in (\ref{simX}) and (\ref{simXplus})
are inconvenient for this
purpose, because extending these to a class of functions
with a variable mode would lead to functions depending non-linearly on the
parameters, which is technically complicated for statistical inference.
For functions that are quadratic in $v_j$,
adding a constant to the argument $v_j$
does not take the function outside the class of quadratic functions;
in other words, horizontal translations can be represented by linear
parameters. Therefore quadratic functions are more useful for our purpose.
It should be noted that quadratic functions are quite common as
representations of choice functions with ideal points that themselves
are explained by other attributes \citep{Jones1983}.

We survey families of quadratic functions with the aim to be
able to represent homophily, attachment conformity, aspiration, as well as sociability.
A model representing pure homophily, the direct analogue to (\ref{simX}), is
\begin{equation}
   a_3(v_j \mid v_i) \,=\, \beta_1 \, (v_j - v_i)^2   \   \label{Q3}
\end{equation}
with $\beta_1 < 0$.
This is too restricted because it forces the optimum value of $v_j$
to be equal to ego's value $v_i$, just like (\ref{simX}).
A more general model proposes separate parameters for each of the four
mechanisms under consideration.
In the following formula we give two equivalent expressions,
the first one showing the linear parametrisation,
the second indicating explicitly the location of the `social norm':
\begin{subequations}\label{Q4}
\begin{align}
   a_4(v_j \mid v_i) & \,=\,  \theta_1 \, (v_j - v_i)^2 \,+\,
   \theta_2\,v_j^2 \,+\, \theta_3\,v_j \,+\,  \theta_4\, v_i \label{Q4a} \\
   & \,\sim \,  \theta_1 \, (v_j - v_i)^2 \,+\,  \label{Q4b}
   \theta_2\,\big ( v_j + \frac{\theta_3}{2\theta_2}\big) ^2 \,+\, \theta_4\, v_i \ ,
\end{align}
\end{subequations}
where the $\sim$ symbol means that the functions differ by only
a constant term, which will be absorbed by the
intercept\footnote{In the SAOM or ERGM representation, the intercept
corresponds to the outdegree effect in the linear predictor.}.

The interpretation can best be based on the three terms in
expression (\ref{Q4b}), and is most straightforward in the case that
$\theta_1$ as well as $\theta_2$ have negative values.
The first term reflects an attraction with a weight
$-\theta_1$ toward $i$'s own value, expressing homophily.
The second term reflects an attraction with a weight $-\theta_2$
toward the value
\begin{equation}\label{norm}
  V^{\rm norm} \,=\, - \frac{\theta_3}{2\,\theta_2} \ ,
\end{equation}
expressing attachment conformity. If this value is within the range
of $V$, it may be regarded as a normative value, and this is
the terminology we shall use.
The third term allows to express a smaller or higher extent
of sociability. In the next section we describe properties of this
social selection function, and the way in which it can express
the four mechanisms.
\medskip

The social selection function (\ref{Q4}) is a quadratic function of the two
variables $v_i$ and $v_j$, and a linear function of
$(v_j - v_i)^2,\ v_j^2, \ v_j$, and $v_i$.
The sender's value $v_i$ and receiver's value $v_j$
are treated differently. As an empirical safety valve it may be advisable
to check whether also an additional free parameter for $v_i^2$
should be included, not directly related to the homophily term;
this leads to an unrestricted quadratic dependence on $v_j$ and $v_i$,
\begin{equation}
   a_5(v_j \mid v_i)  \,=\,  \theta_1 \, (v_j - v_i)^2 \,+\,
   \theta_2\,v_j^2 \,+\, \theta_3\,v_j \,+\,  \theta_4\, v_i
   \,+\, \theta_5\, v_i^2 \label{Q5} \ .
\end{equation}
The interpretations above remain, except that now the tendency to sociability
is expressed by the term $\theta_4\, v_i \,+\, \theta_5\, v_i^2$.

\subsection{Properties of the quadratic representation}

We study some properties of the quadratic functions (\ref{Q4})
and (\ref{Q5}), and elaborate how the four mechanisms are associated
with the four, or five, parameters.

\paragraph{Location of the optimum.}

If $\theta_1 + \theta_2 < 0$, functions (\ref{Q4}) and (\ref{Q5}) are unimodal,
the optimum being located at
\begin{equation}   \label{top}
    v_i^{\textrm{opt}}(\theta) \,=\,
                 \frac{\theta_1 \, v_i \,-\, \theta_3/2 }
                    {\theta_1 \,+\, \theta_2}  \,=\,
                    \frac{\theta_1 \, v_i \,+\, \theta_2 \, V^{\rm norm} }
                   {\theta_1 \,+\, \theta_2} \ ,
\end{equation}
which is a weighted mean of $i$'s own value and the socially normative value.
This could be called the point of attraction, or ideal point,
for an actor with value $v_i$; it is further interpreted below.
If this value is outside the range of $V$, the
location of the optimum must be truncated
and will be assumed at the minimum or maximum value of the range.

\paragraph{Homophily.}

Homophily is expressed directly by the first term in (\ref{Q4b})
and by parameter $-\theta_1$.
The weight for homophily is $\theta_1 / (\theta_1 \,+\, \theta_2)$,
as shown in (\ref{top}).

\paragraph{Attachment conformity.}

Attachment conformity is expressed directly by the second term in (\ref{Q4b})
and its strength by parameter $-\theta_2$.
This term includes two parameters, $\theta_2$ and $\theta_3$.
The weight for attachment conformity  in (\ref{top})
is $\theta_2 / (\theta_1 \,+\, \theta_2)$. The social norm is located
at the value (\ref{norm}), if this is within the range of $V$;
if it is not in this range the attachment conformity has the nature of aspiration,
as discussed next.

The value of (\ref{norm}) can be estimated by plugging in the estimate
$\hat\theta$, depending on the further statistical model used.
Standard errors for $V^{\rm norm}(\hat\theta)$ can then be calculated
using the delta method \citep{Wasserman2004}, see appendix~B.

\paragraph{Aspiration.}

The value of the social norm (\ref{norm}) can be regarded as
a parameter expressing the extent of aspiration\footnote{In this discussion
we only consider positive aspiration;
negative aspiration, an attraction to low values of $V$,
can be treated as its opposite, the directionality
being downward instead of upward.}.
When could one say that variable $V$  has an aspiration aspect?
This may be defined in more than one way, because of the confounding
with homophily. We propose three definitions.
\begin{enumerate}
\item The strongest definition is that, although there may be an element
of homophily, aspiration trumps homophily for everybody,
in the sense that the selection function is increasing on the
entire range of $V$, for every value of $v_i$.
This condition depends on the range of $V$.
Denote the minimum value of $V$ by $V^-$ and its maximum by $V^+$.
For the selection function to be an increasing function of $v_j$
for all $v_i$ in the range of $V$,
given that $\theta_1 < 0, \ \theta_2 < 0$,
the location of the optimum $v_i^{\textrm{opt}}(\theta)$
in (\ref{top}) should be equal to or larger than $V^+$ even for
senders $i$ with $v_i = V^-$. This can be expressed as
\begin{equation}
        \label{aspiration}
        V^{\rm norm}
           \ \geq \  V^+ \,+\, \frac{\theta_1}{\theta_2} \big( V^+ \,-\, V^- \big) \ .
\end{equation}
This can be tested by a right one-sided test of the linear combination
\[
   \theta_3 \,+\,  2 \theta_2\,V^+  \,+\,  2 \theta_1\big( \,V^+ - V^- \big) \ .
\]
  It should be noted that this situation is impossible if $V$ is unbounded
  with $V^+ = \infty$; any quadratic function with $\theta_1 + \theta_2 < 0$
  tends to minus infinity for $v_j \rightarrow \infty$.
  Therefore, the quadratic family proposed here may be less suitable
  for attributes with unbounded range; one possibility to handle this
  is to first transform such attributes to a variable with finite range.
\item  A weaker definition of aspiration is that the contribution
    to the social selection function
    of the terms for the social norm, $\theta_2\, v_j^2 + \theta_3\, v_j$,
    increases in $v_j$.
    This is equivalent to the condition that the location (\ref{top}) of the optimum
    is greater than or equal to the own value $v_i$ for all actors.
    For negative $\theta_2$, this second definition is equivalent to the
    location of the social norm being at least as large as the maximum
    value of $V$, i.e., $ V^{\rm norm} \geq V^+$.
    For positive $\theta_2$, it is equivalent to
    \[
        - \frac{\theta_3}{2\,\theta_2} \leq V^- \ .
    \]
    This can be tested by a right one-sided test of the linear combination
\[
   \theta_3 \,+\,  2 \theta_2\,V^+  \ .
\]
 \item  The weakest definition, in the case that $\theta_2 < 0$,
       is that the location of the norm (\ref{norm}) is larger than
       the mean of $V$.
    This can be tested by a right one-sided test of the linear combination
\[
   \theta_3 \,+\,  2 \theta_2\,\bar V  \ ,
\]
    where $\bar V$ is the mean of $V$. Note that, if $V$ is a centered variable, this is
       equivalent to testing $\theta_3$.
\end{enumerate}
We see that, if $\theta_1 < 0$ and  $\theta_2 < 0$,
the three definitions of aspiration, from strong to weak, are expressed by
progressively weaker lower bounds for $\theta_3$,
which depend on the distribution of $V$.

\paragraph{Sociability.}

Variable $V$ is associated with sociability if higher values of $v_i$
tend to imply that actor $i$, as a sender, has the inclination
to make more tie choices. This means that the social selection function
tends to be higher for higher values of $v_i$.
We propose two definitions.

\begin{enumerate}
  \item A strong definition is that the social selection function
    increases as a function of $v_i$ for all receivers' values $v_j$.
    The derivative of (\ref{Q5}) is
    \begin{equation}\label{derivative}
      \frac{\partial  a(v_j \mid v_i)}{\partial v_i} \,=\,
         2\, (\theta_1 \,+\, \theta_5)\,v_i \,-\, 2\theta_1\,v_j \,+\, \theta_4 \ .
    \end{equation}
    If this value is non-negative for all values $v_i, v_j$,
    the strong definition is satisfied. Depending on the signs of the
    coefficients, only one of the combinations of $V^-, V^+$ needs to be checked.
    In the case that $\theta_1 \leq 0,\ \theta_1 + \theta_5 \leq 0$,
    the condition is
    \[
         2\, (\theta_1 \,+\, \theta_5)\,V^+ \,-\, 2\theta_1\,V^- \,+\, \theta_4 \geq 0 \ .
    \]
  \item A weak definition for $V$ to have a sociability dimension
        is that the optimum value
        of the social selection function for given $v_i$,
\begin{equation}
        \label{thetop}
        a^{\rm opt}(v_i) \,=\, \max_{v_j}  a(v_j \mid v_i) \ ,
\end{equation}
    increases with $v_i$.
    If $\theta_1 + \theta_2 < 0$ , the optimum is assumed for
    $v_j =  v_i^{\textrm{opt}}(\theta)$ given in (\ref{top}).
    Some calculations show that the value of the optimum is
\begin{equation}
        \label{themax}
    a^{\rm opt}(v_i) \,=\,
        \frac{2\,\theta_1 \, \theta_2}{\theta_1 \,+\, \theta_2}
                \big(v_i - V^{\rm norm}\big)^2
        \,+\, \theta_4\, v_i \,+\, \theta_5 \, v_i^2 \ .
\end{equation}
    This is an increasing function of $v_i$ if
    \begin{equation}\label{sociable}
     \frac{4\,\theta_1 \, \theta_2}{(\theta_1 \,+\, \theta_2)^2}
      \big(v_i - V^{\rm norm}\big)
        \,+\, \theta_4 \,+\, 2 \theta_5 \, v_i \geq 0 \ .
    \end{equation}
    Since the latter function is linear in $v_i$, it needs to be checked only
    for the extremes $v_i = V^-, V^+$.

    It is possible that (\ref{top}) is outside of the range of $V$
    for some values $v_i$. Then for such values, the optimum
    is assumed at the minimum or maximum value of the range,
    and the value of the optimum has to be calculated accordingly.
\end{enumerate}
In many cases a visual check of the plotted function will easily show whether the weak or
strong versions of sociability are satisfied. The formulae show that
the strong version depends on parameters $\theta_1, \ \theta_4$, and $\theta_5$.
The weak version depends on all five parameters.
But always, higher values of $\theta_4$ are conducive to the association
of $V$ with sociability.
\medskip

These interpretations seem reasonable but may not always be compelling.
They are based, theoretically, on the assumption that attraction for sending ties,
as far as dependent on the values of $V$, can be expressed as a combination of homophily
and attraction to a common normative value (which may be a hypothetical value
outside of the range of the possible); and, empirically, on the fit of the quadratic
shape of the social selection function in whatever statistical network model is
being utilized.
Further, they ignore other elements of the model specification,
which can depend on variables or network positions that may be associated
to $V$ in some way.

\smallskip

Concluding, this reasoning leads to a four-parameter quadratic model
in $v_i$ and $v_j$, which may be extended to a five-parameter model.
Quadratic social selection functions have been used occasionally
in statistical network modeling.
Examples are \citet[][eq. (16)]{REP2001}, where a quadratic selection function is
mentioned as a possibility for non-directed networks, without elaboration
or example; and \citet{MerckenCSHM2012}, using a squared term
of alter's smoking habits in a co-evolution study of friendship and smoking.
Our proposal is to use them more systematically.

\section{Example}
\label{S_ex}

We demonstrate the empirical value of this approach by analysing
a longitudinal network
of advice ties among students enrolled in a master degree program in
business administration (MBA), with academic performance and age as the
actor attributes under consideration.
The data were collected by Vanina Torl\`{o}.
The network was composed of full-time students in an elite Italian school
for professional management education.
This network was analyzed earlier in \citet{SLT2013}.
Educational settings provide an ideal context for the study of
homophily-related network processes because -- contrasting with behavior
in the context of formal organizations --
students' behavior is hardly affected by pre-assigned roles
or by differences in formal hierarchical positions.
The cohort consisted of 75 students, providing full response for all variables.
The program had a duration of one year, and data collection for the
three panel waves took place close to examination periods in
March, July, and November.

\begin{figure}[htb]
\includegraphics[scale=0.22]{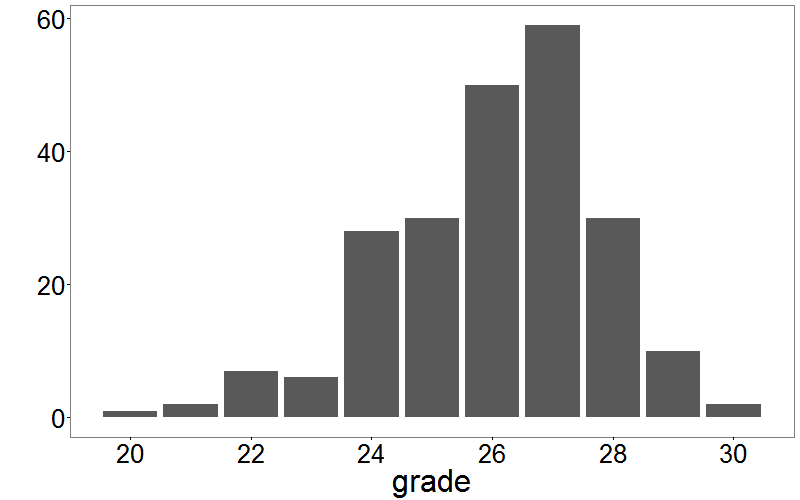}
\hfill
\includegraphics[scale=0.22]{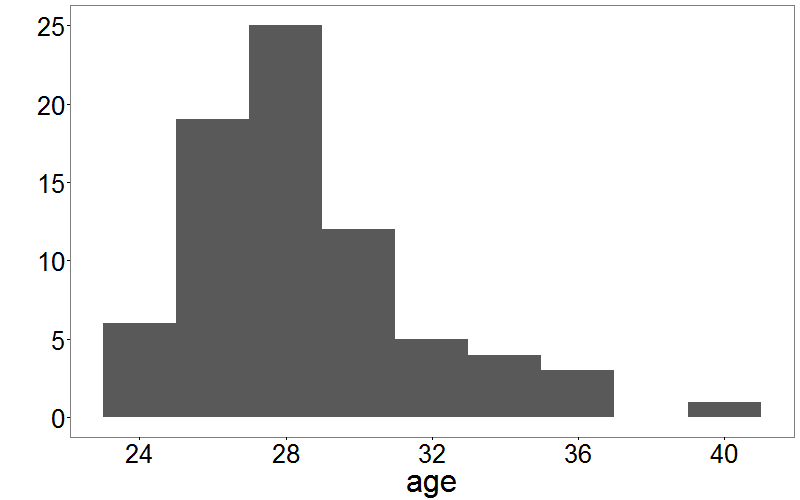}
  \caption{Left: bar plot of grades, pooling the three waves; right: histogram of ages.
  Frequencies on vertical axis.}
  \label{F1}
\end{figure}

For the advice relation, respondents were asked to indicate the names of other
students whom they regularly consulted for help and support on
program-related tasks; examples mentioned were asking for class notes,
help in solving homework problems, etc. Any number of classmates
could be mentioned.
Academic performance was measured as the average grade,
rounded to integers, on the 10-12 exams in the examination period,
calculated from information supplied by the MBA office.
The range of academic grades was 20--30, with a mean of 26.
Age ranges from 24 to 40 years, with an average of 29 years.
The distributions of grades and age are shown in Figure~\ref{F1}.
Further information on this data set can be found in \citet{LSST11}.

Academic grades are important in the context of professional management
education, and may be expected to be important for structuring
interpersonal advice relations \citep{LSST11}.
In the extremely competitive context of an MBA class,
academic performance, as represented by grades, is treated as a signal
of students' commitment, sense of duty, and competence ---qualities
valued both by potential employers as well as potential business partners.
A strong emphasis is therefore placed on performance when it comes to
forming a student's network partners for getting advice in
academic matters.
This suggests that grades could have an aspiration dimension for the
advice relation. However, asking advice is much easier in dyads with
high mutual social acceptance.
The tension between the two objectives of individual achievement and
social acceptance has indeed been a central factor in the economics and
sociology of schooling at least since the ``Coleman Report''
\citep{Coleman1966} and perhaps even earlier \citep{Coleman1961}.
The best performers may become quite selective and reciprocate
ties only to those with a similarly high performance; in anticipation,
lower performers may be reluctant to ask their advice. In addition,
social acceptance may be higher between students of similar performance.
This would lead to grade-related homophily in advice. If social acceptance
would generally be higher for students of normative performance (whatever
the normative level is, provided it is less than
the maximum possible value), there would be a attachment conformity dimension.
For the sociability dimension, one may argue in two opposite ways:
those with high grades are less in need of advice,
so they will ask less for advice, i.e., grades represents a negative sociability
mechanism;
or those with high grades find academic performance more important,
therefore are more active generally also in advice asking,
reflected by positive sociability.
This shows that a priori, all four mechanisms of aspiration, homophily,
attachment conformity, and sociability might be associated with grades in their effect
on advice asking.

Age is expected to be of lower importance than grades for selection
of advisors. Homophily still might be relevant, but rather as
a consequence of general social interaction being easier between
students of similar age than as a mechanism related to advice
specifically. Formulating prior expectations would be quite speculative;
therefore we refrain from doing so.
Age is included here to improve the fit of the model and to have a second
illustration of a numerical actor variable, with possibly a different,
less important role for structuring the advice network.

We estimate the Stochastic Actor-oriented Model (SAOM) for this data set.
Explanations and further background to this model
are given in \citet{SnijdersEA10b} and \citet{Snijders2017}.
Single parameters are tested by $t$-tests (dividing parameter
by standard error and testing in a standard normal distribution);
multi-dimensional tests are tested by Wald-type tests with
a chi-squared null distribution \citep{SienaManual18}.
This is supported not by mathematical proofs but by numerous
simulation studies.
Significance will be gauged at the conventional level of $\alpha=0.05$.
For the analysis we used the R package RSiena \citep{SienaManual18},
version~1.2-8.
The implementation of the five-parameter model in RSiena is briefly
treated in Appendix~A.

\subsection{Results}

We present results for a model that includes the five effects
(\ref{Q5}) of grade and age.
Both variables are centered.
The structural part of the model is defined
in a way that is now more or less standard for stochastic actor-oriented models
\citep{SnijdersEA10b,SienaManual18} with reciprocity,
three degree-related effects to reflect the variances and
correlations of degrees, and transitivity implemented by the
Geometrically Weighted Edgewise Shared Partners
(`gwesp') statistic \citep{SPRH06,Hunter2007}.
This gives here a better fit than the more traditional
specification by a count of transitive triplets. Also an interaction
between reciprocity and transitivity is included \citep[cf.][]{Block2015};
even though not significant, this effect improved the goodness of fit
of the model.
Additional homophily for categorical actor variables
is included in the model for gender and the binary variable nationality
(Italian vs.\ non-Italian).

Parameter estimates are presented in Table~\ref{T1}.
The tables give asterisks according to two-sided tests, but for parameters
$\theta_1$ and $\theta_2$ we can have in the back of our mind
that they are expected to be negative.
Goodness of fit for distributions of indegrees, outdegrees,
and geodesic distances, as well as for the triad count, was
tested by the \texttt{sienaGOF} function \citep{SienaManual18}.
This operates by simulating networks according to the model with
estimated parameters, and comparing the observed values of
selected statistics with their distributions in the simulated set
of networks. The fit is then assessed by comparing the Mahalanobis distance
of the observations to the mean of the simulated values and computing
the associated $p$-value. For all four sets of statistics mentioned
the $p$-value was between 0.10 and 0.90, which means the fit was good.

\begin{table}[htb]
  \centering
{\small
\begin{tabular}{l r@{.}l r@{.}l  }
\hline
\rule{0pt}{2ex}\relax
Effect &\multicolumn{2}{c}{par. \ }&   (s&e.)  \\[0.5ex]
\hline
\rule{0pt}{2ex}\relax
Rate period 1 & 7 & 939 & (0 & 691)\\
Rate period 2 & 5 & 883 & (0 & 471)\\
\hline
\rule{0pt}{2ex}\relax
outdegree                                  & --2 & 181$^{\ast\ast\ast}$ & \hspace*{2em} (0 & 208)\\
reciprocity                                &   1 & 606$^{\ast\ast\ast}$ & (0 & 197)\\
transitivity gwesp                         &   1 & 307$^{\ast\ast\ast}$ & (0 & 121)\\
reciprocity $\times$ transitivity gwesp \hspace*{2em}
                                           & --0 & 314                  & (0 & 250)\\
indegree -- popularity                     &   0 & 0253$^{\ast\ast}$    & (0 & 0089)\\
outdegree -- popularity                    & --0 & 101$^{\ast\ast}$     & (0 & 033)\\
outdegree -- activity                      & --0 & 0072                 & (0 & 0092)\\
gender alter (M)                           &   0 & 027                  & (0 & 098)\\
gender ego (M)                             & --0 & 239$^\ast$           & (0 & 100)\\
same gender                                &   0 & 130                  & (0 & 092)\\
same nationality                           &   0 & 405$^{\ast\ast\ast}$ & (0 & 122)\\
$\hat\theta^{\rm{g}}_1$ (grades ego minus alter) squared  & --0 & 0288$^{\ast\ast\ast}$     & (0 & 0073)\\
$\hat\theta^{\rm{g}}_2$ grades squared alter              & --0 & 003              & (0 & 012)\\
$\hat\theta^{\rm{g}}_3$ grades alter                      &   0 & 044              & (0 & 032)\\
$\hat\theta^{\rm{g}}_4$ grades ego                        & --0 & 095$^{\ast\ast}$ & (0 & 031)\\
$\hat\theta^{\rm{g}}_5$ grades squared ego                &   0 & 026$^{\ast\ast}$ & (0 & 010)\\
$\hat\theta^{\rm{a}}_1$ (age ego minus alter) squared     & --0 & 0014             & (0 & 0023)\\
$\hat\theta^{\rm{a}}_2$ age squared alter                 & --0 & 0070             & (0 & 0045)\\
$\hat\theta^{\rm{a}}_3$ age alter                         &   0 & 039$^\ast$       & (0 & 019)\\
$\hat\theta^{\rm{a}}_4$ age ego                           &   0 & 038$^\ast$       & (0 & 018)\\
$\hat\theta^{\rm{a}}_5$ age squared ego                   & --0 & 0071$^\dagger$   & (0 & 0041)\\
\hline
\multicolumn{5}{l}{\footnotesize{$^\dagger$ $p$ $<$ 0.10;
$^\ast$ $p$ $<$ 0.05; $^{\ast\ast}$ $p$ $<$ 0.01;
$^{\ast\ast\ast}$ $p$ $<$ 0.001 (two-sided).}}\\
\multicolumn{5}{l}
   {\footnotesize{convergence $t$ ratios all $<$ 0.04;
      overall maximum convergence ratio 0.09.}}\\
\multicolumn{5}{l}
{\footnotesize{Number of decimals presented depends on standard errors.}}\\
\vspace*{0.01em}
\end{tabular}
}
  \caption{Parameter estimates and standard errors for the advice network
  between MBA students. Grades and age are centered.}\label{T1}
\end{table}


Before discussing the effects of grade and age, we give a very brief
discussion of the other effects (all under the usual
`if everything else is equal' clause). There are the expected
strong reciprocity  and transitivity effects.
The positive indegree-popularity and negative
outdegree-popularity effects show that advice is asked with higher
probability from those who already give much advice, and those who ask little for it;
this corresponds to the nature of advice giving.
Males tend to ask for advice less than females, and advice is asked more
from students having the same nationality. Thus, there is evidence for
homophily with respect to nationality, but the table shows
this is not significantly the case for gender.

The effect of grades on advice is important. The joint test of the five
parameters yields $\chi^2_5 = 23.3,\ p < 0.0005$.
The effect of age is also significant, but less strongly so,
$\chi^2_5 = 11.9,\ p < 0.05$.
Testing whether the quadratic effects are an improvement on model~(\ref{egoXaltX})
without any quadratic terms, we find that the two quadratic effects
of grades are jointly significant  ($\chi^2_2 = 6.6,\ p < 0.05$), and those
for age likewise ($\chi^2_2 = 10.1,\ p < 0.01$).
For both variables, the squared ego term is significant at $p < 0.10$,
so model~(\ref{Q5}) seems indeed to be slightly better than model~(\ref{Q4}).
Summarizing, there is strong evidence for influence of grades on the advice network,
as well as evidence for influence of age.
This confirms the applicability of the five-parameter model
to this data set, for grades and also for age.

Over and above issues of model fit,
the five-parameter model also affords interpretation
of the effects of grades and age on performance.
The social selection function for grades can be interpreted in the following way.
The centered grades variable, i.e., grades~--~26.1,
is denoted by $V$. This ranges from $V^- = -6$ to $V^+ = 4$.

\begin{figure}[htb]
  \centering
\includegraphics[scale=0.3]{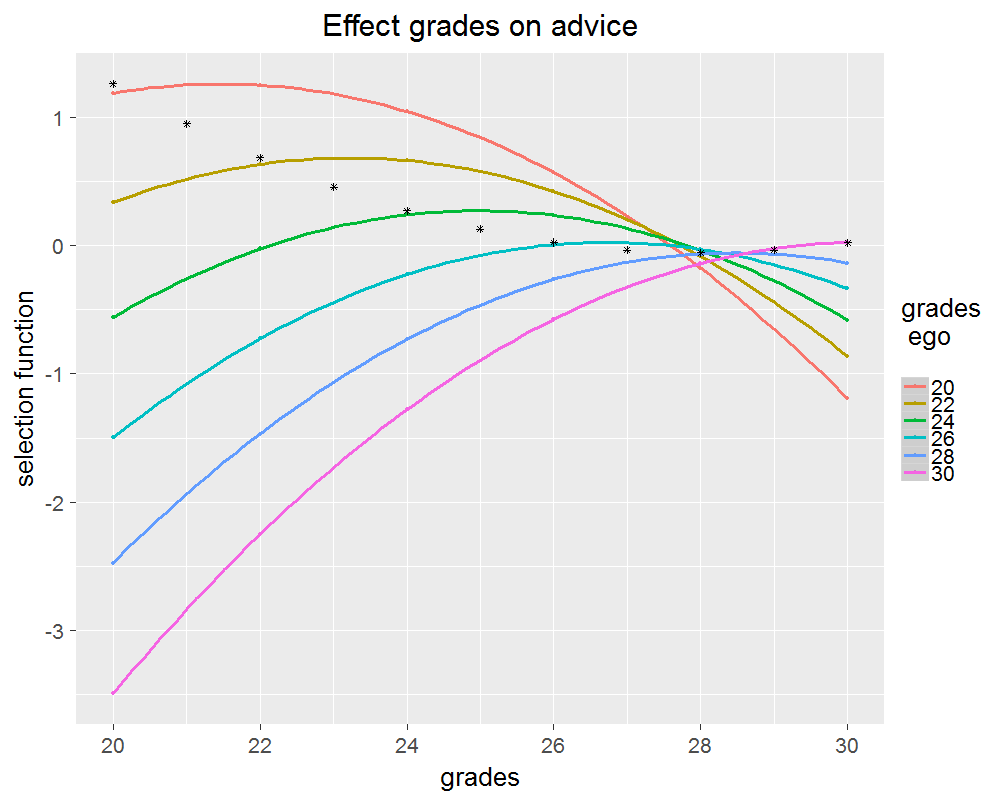}
  \caption{Social selection function for the effect of grades on advice
  for the model in Table~\ref{T1}.
  The continuous curves are the social selection function,
  separately for six values of ego's grade from 20 to 30, as a function of
  alter's grade (horizontal axis). The asterisks indicate the
  maximum of the social selection function as a function of
  ego's grade on the horizontal axis.
  }
  \label{F2}
\end{figure}

\begin{enumerate}
  \item There is a clear and strongly significant
        aspect of homophily, with $\hat\theta_1 = - 0.0288 < 0$.
  \item The coefficient of grades squared,  $\hat\theta_2 = - 0.003$, is negative,
        so we can elaborate the potential interpretation of an attraction to a
        socially normative value. However, the parameter is not
        significantly different from 0, so the interpretation is not very strong.\\
        The estimated value of the social norm (\ref{norm}) is
        $\hat V^{\rm norm} \,=\, - \hat\theta_3 /(2\,\hat\theta_2)  \,=\, 6.9$,
        higher than the maximum value of $V$.
        Therefore, the second definition of aspiration is satisfied
        with respect to the parameter estimates, although there
        is no statistical significance to support this.
        The weight for attachment conformity is only
        $\hat\theta_2/(\hat\theta_1 + \hat\theta_2) = 0.1$
        while it is 0.9 for homophily.
        In other words, homophily dominates attachment conformity.\\
        Because the coefficient $\hat\theta_2$ is far from significant,
        it is not meaningful to calculate a standard error for $\hat V^{\rm norm}$.
  \item The social selection function is plotted in Figure~\ref{F2}.
        For egos with low grades, it is almost equally strongly decreasing
        as it is increasing for egos with high grades; this is in line
        with the low weight for attachment conformity.
 \item  Sociability, represented by the optimum (\ref{thetop}) of the social selection
        function, is plotted by the asterisks in Figure~\ref{F2}.
        It is decreasing for the lower half of the range of grades, and
        approximately constant for the upper half. Although not
        decreasing uniformly, this plot nevertheless suggests a weakly negative
        sociability aspect for grades ---but weaker even than the weak definition.
\end{enumerate}

For age, similarly, the social selection function can be studied.
Now $V$ is age in years, of which the mean of 29 years is subtracted.
It ranges from --5 to +11.

\begin{figure}[htb]
  \centering
\includegraphics[scale=0.28]{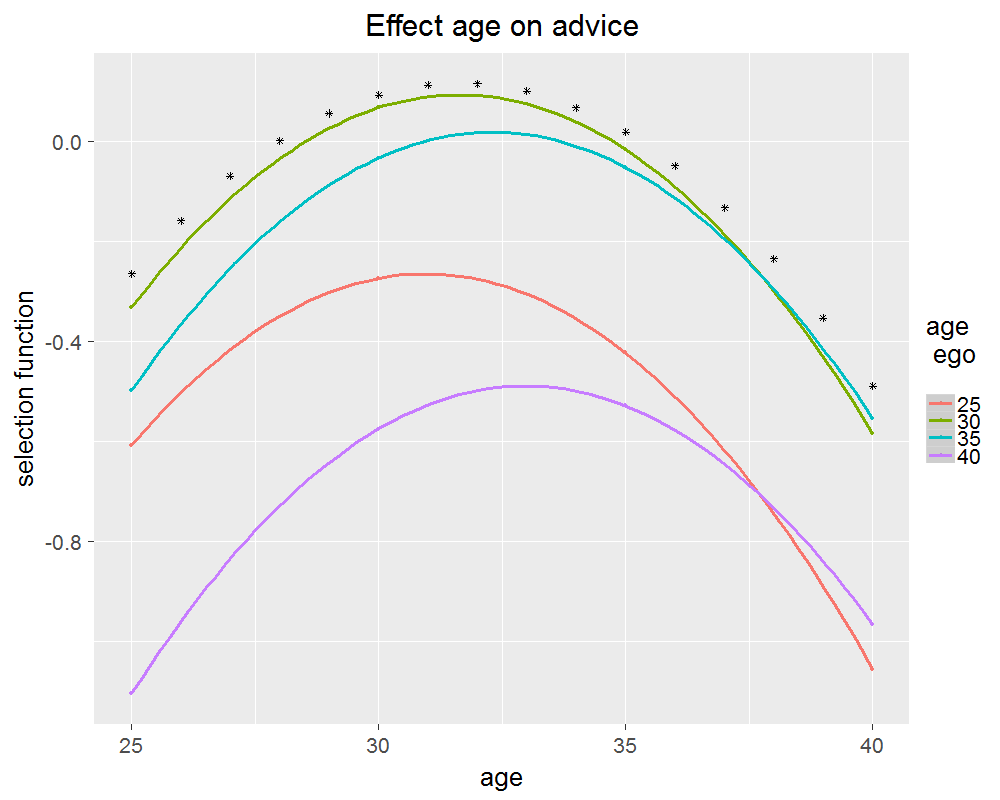}
  \caption{Social selection functions for the effect of age on advice
  for the model in Table~\ref{T1}.
  The continuous curves are the social selection functions,
  separately for four values of ego's age from 25 to 40 years, as a function of
  alter's age (horizontal axis). The asterisks indicate the
  maximum of the social selection function as a function of
  ego's age on the horizontal axis.
  }
  \label{F2_age}
\end{figure}

\begin{enumerate}
  \item Both $\hat\theta_1$ and $\hat\theta_2$ are negative; the former is clearly
    not significant ($\hat\theta_1 = - 0.0014$, $\text{s.e.} = 0.0023$),
    the latter has $p < 0.10$ ($\hat\theta_2 = - 0.0070$, $\text{s.e.} = 0.0045$),
    and is significant in a one-sided test. This shows that for age, the aspect
    of homophily is not significant and there is weak support for an aspect of attachment conformity.
  \item The estimated value of the social norm (\ref{norm}) is
        $\hat V^{\rm norm} \,=\, - \hat\theta_3 /(2\,\hat\theta_2)  \,=\, 2.8$,
        corresponding to 32 years, and higher than the mean age.
        Therefore, the last definition of aspiration is satisfied.
        Thus, there is a weak aspiration aspect; this is significant,
        as $\hat\theta_3$ is significantly positive; however, only
        its weakest definition is satisfied,
        and the social norm is not much higher than the mean age.
  \item The social selection function is plotted in Figure~\ref{F2_age}.
        We see that the location of the optimum hardly changes
        with ego's age. It is noteworthy that the range
        (i.e., maximum minus minimum) of the selection function is~1.3,
        much less than the range of about~5
        of the social selection function for grades, which underscores that
        age is much less important than grades for the advice relation.
 \item  The main difference between the social selection functions for different values
        of ego's age is that it is highest for egos of medium age,
        and lower for egos who are on the young or on the old side.
        This is exhibited by the dashed line in Figure~\ref{F2_age},
        giving the value of the optimum depending on ego's age.
\end{enumerate}

Summarizing, for age there is no homophily and a weak attachment conformity dimension, with a
slight aspiration aspect; and sociability is highest for medium values of age.

\paragraph{Comparison with other social selection functions for grades}

\begin{figure}[htb]
  \centering
\includegraphics[scale=0.175]{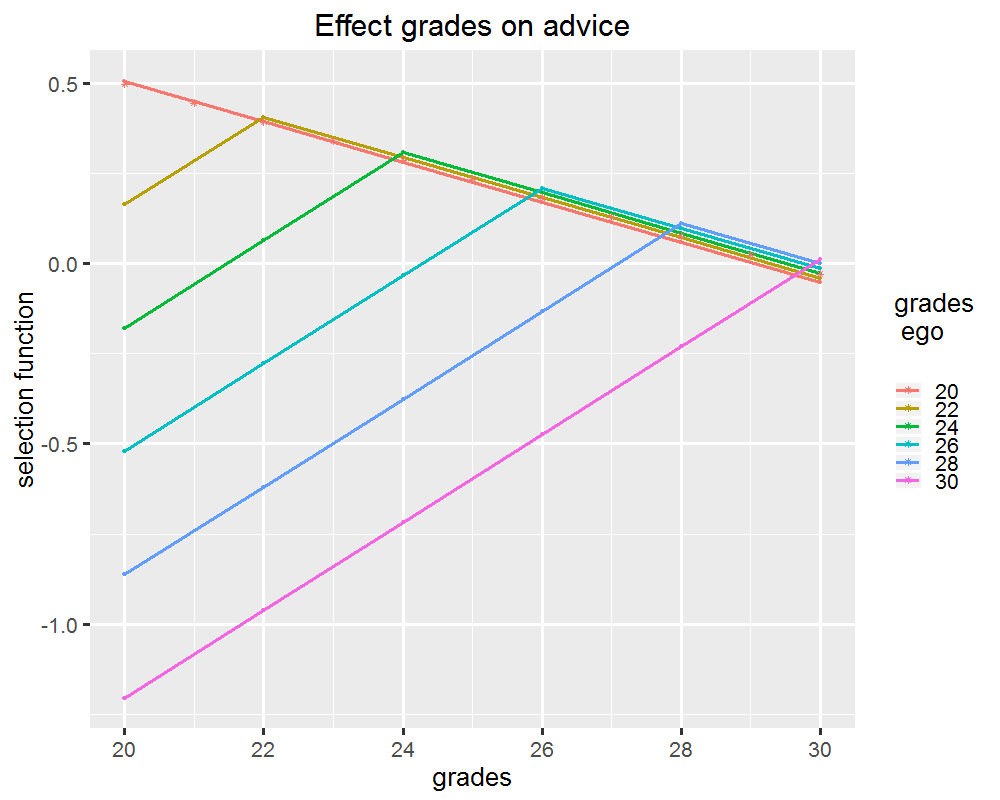}
\hfill
\includegraphics[scale=0.175]{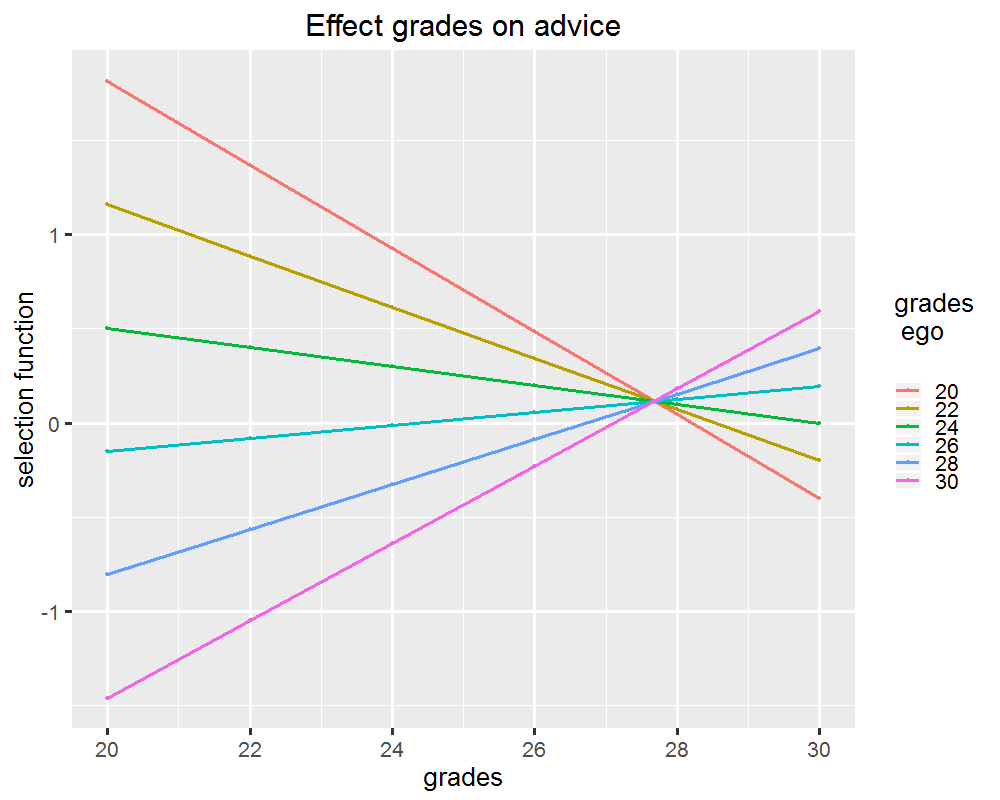}


  \caption{Social selection function for advice depending on
  ego's and alter' grades.
  Left: similarity specification with main effects
  (\ref{simXplus}); right: linear interaction specification (\ref{egoXaltX}).}
  \label{F2ea.si}
\end{figure}

Models with the social selection functions defined by the
main effects for ego and for alter with the
absolute differences (\ref{simXplus}) and
with ego-by-alter interaction~(\ref{egoXaltX}) were also estimated.
Model~(\ref{egoXaltX}) fitted less well than (\ref{Q4}).
Plots are presented in Figure~\ref{F2ea.si}.
The purpose of the figures is to demonstrate the consequences of
the model assumptions.
The differences with Figure~\ref{F2} show that
the representation of this data set by the quadratic model
is quite different from the representation using the absolute
difference or the product interaction.
For model (\ref{simXplus}) the social selection functions for
$v_j > v_i$ are almost on the same line, which is the case because
parameters $\beta_1$ and $\beta_3$ cancel each other almost precisely
for $v_j > v_i$.
When looking carefully it turns out that the lines in
the right-hand side figure are not so strongly different from the
curves in Figure~\ref{F2}, and indeed the fit for Model~(\ref{egoXaltX})
is not much worse than the fit for Model~(\ref{Q4}).

\section{Summary and discussion}

This paper is about how to specify effects of numerical actor attributes,
satisfying an interval level of measurement,
in statistical models for directed social networks, where the set of network ties
constitutes the dependent variable.
Effects of actor attributes on networks are not as straightforward as main
effects in a generalized linear model, because the dependent variable is
defined at the dyadic level: ordered pairs of actors; whereas the attribute is
defined at the monadic level: actors. Some transformation
from the actor level to the dyadic level is necessary.

This paper considers directed networks where a tie from
sender $i$ to receiver $j$ can be interpreted as the result of a positive
choice, in some sense, originating from $i$ to the target $j$.
This allows us to interpret the effects of the attribute as a way
of structuring attraction between actors.
Homophily is a major mechanism of attraction, as discussed by
\citet{LazarsfeldMerton54} and many others, cf.\ \citet{McPhersonEA01}.
While homophily is often a mechanism of primary importance,
it may be not always the strongest and it can easily be
confounded with other rival mechanisms.
Our model combines a diversity of mechanisms:
homophily, i.e., attraction to similar others (also called assortativity);
aspiration, attraction to high values;
attachment conformity, attraction to a value common for all in the network,
which might be called a normative value;
and sociability, the inclination to make many tie choices.
Each of these mechanisms could be associated to a larger
or smaller extent with the actor attribute in question.
Choices by social actors are likely to be steered by multiple considerations,
hence these mechanisms may well be confounded.

The mathematical specification of our model is a quadratic function (\ref{Q4}),
extendable to (\ref{Q5}),
of the attribute values of the sender and the receiver of the tie.
This function can be used in a linear predictor
in any statistical network model; our example was for a
Stochastic Actor-oriented Model, but our reasoning applies likewise
to other statistical models, e.g., the Exponential Random Graph Model
\citep{WassermanPattison96,ERGMbook}.
For the Stochastic Actor-oriented Model, Appendix~A mentions the effects
that can be used for implementing (\ref{Q4}) and (\ref{Q5}).

For the interpretation of the model, considering the figure
is the best option. The four or five parameters separately are hard
to interpret, because their effects are combined.
Nevertheless something can be said about the association between the four
mechanisms and the parameters in the statistical model.
For homophily this is straightforward, it is
represented by a single parameter (i.e., by $-\theta_1$).
Attachment conformity and aspiration are inseparable in the model because
both are associated  with the location of the social norm.
Aspiration means conformity to the notion that ties should be sent
especially to actors with a high value of the attribute.
Together these mechanisms are represented by two parameters
(i.e., $-\theta_2$ and  $\theta_3$).
Sociability has the least direct interpretation in terms of
model parameters, and to interpret it the figure will be required.
\medskip

The quadratic family proposed in this paper is only a relatively simple version
of the realm of possibilities.
Other monadic-to-dyadic transformations could be used
to operationalize the combination of the four confounded mechanisms.
This choice should be based on considerations of theory and empirical fit.
One possibility is to use model (\ref{Q4}) but replace the
squared difference between ego and alter by the absolute difference.
This model can be obtained also as an extension of (\ref{simXplus}).
Whether the kink in the function at $v_j = v_i$ is an advantage
or disadvantage will depend on the research in question.

Other possible transformation are cubic and higher-order polynomials,
which will yield more flexibility but also increase the number of parameters.
Splines or fractional polynomials could also be considered
\citep[e.g.,][]{SauerbreiEA2007}.
For example, a cubic function could represent that the selection curves
may be wider for high than for low values of~$V$.
For actor attributes with an unbounded range,
the quadratic transformation and other polynomials may have a
less good fit especially at for extreme values, because they tend
to positive or negative infinity unless the function is exactly constant;
this does not necessarily hold for fractional polynomials.

Other extensions of these models are possible by proposing interactions of
these variable-related mechanisms with structural effects such as reciprocity,
endogenous popularity, and transitivity. For example, reciprocity
could attenuate a tendency toward homophily as argued by
\citet{Block2018}. There may be arguments, theoretical and/or empirical,
for other interactions between the four mechanisms treated in this paper
with structural network effects, and this is an interesting topic
worth of further study.

With respect to what is outside of our scope conditions the following can be said.
For non-directed networks the arguments based on regarding the tie
as a directional choice by the sender do not apply. Quadratic transformations of
numerical variables may be useful there, too; but we do not go into
discussing interpretations of such models. For dichotomous variables there
are only three degrees of freedom, which are included in models
(\ref{simXplus}) and (\ref{egoXaltX}), and the quadratic models are
superfluous. For categorical variables the situation is more complex,
and we do not feel able to propose any generalizable ideas in this paper.

An important caveat for interpretations is that,
just like in other generalized linear models, these attribute effects are
net of the further effects included in the network model,
and the other effects may be correlated in complex ways
with the attribute effects.
This implies that the social selection function cannot be interpreted
as something akin to a true preference function, although the terms we have been
using may suggest this.
For example, we do not regard the normative value  $V^{\rm norm}$
as a revealed norm in any real sense.

For the example presented here, in
the analysis of the evolution of an advice network in an MBA cohort
with academic grades as the salient attribute,
there was clear evidence of homophily,
and the medium-strength definition of aspiration was satisfied,
although not significantly.
There were no signs of attachment conformity other than aspiration,
or of sociability associated with grades.

To conclude, the four-parameter model (\ref{Q4})
is attractive theoretically and can `let the data speak for themselves'
about how the elements of homophily, aspiration, and attachment conformity may combine
in any specific empirical setting.
The five-parameter model (\ref{Q5}) is a more flexible version that will sometimes
be an improvement.
We propose that empirical researchers have these models in mind
when estimating statistical models for directed networks with
numerical actor attributes, and we expect that in many cases
these specifications will be appropriate.
This does not imply that we suggest to necessarily
use such quadratic models for all numerical actor variables.
They are less parsimonious, with four or five parameters instead of only one for
the absolute difference model (\ref{simX}) and three for
models (\ref{simXplus}) and (\ref{egoXaltX}).
For cases where the dependence of the network on the attribute
is weaker, it may be found that one of these three models provides
a good enough approximation, so that the conclusions reached
will be basically the same and the goodness of fit still adequate.
Then for the empirical analysis the quadratic model is not necessary,
although theoretically it may still be preferable.

\section*{Appendix A. Implementation in RSiena}

Models (\ref{Q4}) and (\ref{Q5}) can be implemented in the
RSiena package \citep{SienaManual18} by the following effects.
The `shortNames' are the shorthand codes that can be used
to specify the effects in RSiena.
\medskip

\begin{center}
 \begin{tabular}{lll}
  name & shortName  & \phantom{$\Sigma$} $s_{ki}(x,v)$ \\
 \hline
$V$ \  ego                     & \texttt{egoX}     &  $ \sum_j x_{ij}\, v_i         $    \\
$V$ \  ego minus alter squared & \texttt{diffSqX}  &  $ \sum_j x_{ij}\, (v_i-v_j)^2 $   \\
$V$ \  alter                   & \texttt{altX}     &  $ \sum_j x_{ij}\, v_j         $     \\
$V$ \  alter squared           & \texttt{altSqX}   &  $ \sum_j x_{ij}\, v_j^2       $      \\
$V$ \  ego $\times$  alter & \texttt{egoXaltX} &  $ \sum_j x_{ij}\, v_i\,v_j $   \ .  \\
 \end{tabular}
\end{center}
\medskip

\noindent
All five effects are simple transformations from the monadic to the dyadic level,
so that the implementation will be straightforward also for other statistical
network models and other software, using calculated dyadic covariates.

\section*{Appendix B. Standard errors}

The location (\ref{norm}) of the social norm,
$V^{\rm norm}(\theta) $, is a nonlinear function of the parameter vector $\theta$.
When $\theta$ is estimated by some estimator $\hat\theta$,
the standard error of $V^{\rm norm}(\hat\theta)$ can be
obtained from the delta method \citep{Wasserman2004}.
According to this method, an approximation
to the covariance matrix of a function $f(Z)$ of a random vector $Z$
is given by
\begin{equation}
   \Cov\big(f(Z)\big) \,\approx\,   D' \,\Cov(Z)\,D
\end{equation}
where $D$ is the gradient
\[
   D \,=\, \frac{\partial f(z)}{\partial z}  \Bigg\rvert_{\displaystyle z=\mbox{E}(Z)} \ .
\]
This is applied to $Z = \big(\hat\theta_2, \hat\theta_3 \big)$
with the function (\ref{norm})
\[
   f(\theta_2, \theta_3) \,=\,  \frac{\theta_3}{-2\theta_2}
\]
and
\[
 D \,=\, \left(
          \displaystyle \frac{\theta_3}{2\theta_2^2} , \hspace{0.7em}
          \displaystyle  \frac{-1}{2\theta_2}
           \right) \ ,
\]
filling in the estimate for the expected value of $\hat\theta$.
For  Method of Moments estimation in the Stochastic Actor-oriented Model,
$\Cov(\hat\theta)$ is obtained as in \citet{Snijders01}.

\section*{Acknowledgements}

We thank Vanina Torl\`{o} for permission to use her data set.
%



\end{document}